\documentclass[aps,prl,reprint,superscriptaddress,floatfix]{revtex4-2}

\usepackage{graphicx}
\usepackage{physics}
\usepackage{xcolor}
\usepackage[section]{placeins}
\usepackage{amsmath}
\usepackage{bm}
\usepackage{braket}
\usepackage[hidelinks]{hyperref}

\begin{document}

%Title of paper

\title{Ultrafast cooperative electronic, structural, and magnetic transition in $\alpha$-MnTe}

\normalsize

\author{Tiangao Lu}
\thanks{These authors contributed equally to this work}
\affiliation{Department of Physics, National University of Singapore, Singapore 117551, Singapore}
\affiliation{Center for Quantum Technologies, National University of Singapore, Singapore 117543, Singapore}

\author{Ao Wu}
\thanks{These authors contributed equally to this work}
\affiliation{Department of Physics, National University of Singapore, Singapore 117551, Singapore}

\author{Junxiang Li}
\thanks{These authors contributed equally to this work}
\affiliation{Department of Physics, National University of Singapore, Singapore 117551, Singapore}

\author{Qiyuan Wu}
\affiliation{Department of Physics, National University of Singapore, Singapore 117551, Singapore}

\author{Meng Zeng}
\affiliation{Department of Physics, State Key Laboratory of Quantum Functional Materials, and Guangdong Basic Research Center of Excellence for Quantum Science, Southern University of Science and Technology (SUSTech), Shenzhen, Guangdong 518055, China}

\author{Di Cheng}
\affiliation{Department of Physics, National University of Singapore, Singapore 117551, Singapore}

\author{Chang Liu}
\affiliation{Department of Physics, State Key Laboratory of Quantum Functional Materials, and Guangdong Basic Research Center of Excellence for Quantum Science, Southern University of Science and Technology (SUSTech), Shenzhen, Guangdong 518055, China}

\author{Jiangbin Gong}
\affiliation{Department of Physics, National University of Singapore, Singapore 117551, Singapore}
\affiliation{Center for Quantum Technologies, National University of Singapore, Singapore 117543, Singapore}

\author{Xinwei Li}
\email[email: ]{xinweili@nus.edu.sg}
\affiliation{Department of Physics, National University of Singapore, Singapore 117551, Singapore}

\date{\today}
	
\begin{abstract}

Femtosecond laser control of antiferromagnetic order is a cornerstone for future memory and logic devices operating at terahertz clock rates. Here, we demonstrate an ultrafast laser-induced phase transition in $\alpha$-MnTe that simultaneously engages charge, lattice, and spin degrees of freedom. Using time-resolved reflectivity and birefringence measurements, we reveal that photo-induced demagnetization is accompanied by pronounced structural and electronic instabilities, manifesting as a multistage temporal evolution following photoexcitation. The transition uncovers unique spin-charge and spin-lattice coupling pathways, which we discuss in the context of recently proposed altermagnetic phenomenology.

\end{abstract}

\maketitle

The study of antiferromagnetic spintronics addresses the growing demand for memory and computing units that combine fast processing speeds, resilience to charge perturbations, and compact footprints \cite{Baltz2018,Jungwirth2016}. As the push toward terahertz clock rates intensifies, a key scientific and technological objective is to realize laser control of antiferromagnetic order on femto- to picosecond timescales. The past decade has seen remarkable progress on this front, with promising demonstrations including the coherent manipulation of magnons \cite{Kampfrath2011,Lu2017,Zhang2024,Leenders2024}, phonon-driven metastable spin states \cite{Disa2020,Afanasiev2021,Ilyas2024}, and photo-induced magnetic phase transitions \cite{Li2013,Kurihara2018,Schlauderer2019,Fichera2025}. 

A cornerstone of all studies on laser-controlled magnetism is to achieve strong interaction between light and spin -- two entities that are typically only weakly coupled directly but can be bridged via intermediaries involving other fundamental degrees of freedom such as charge \cite{Li2013,Li2025} or lattice \cite{Mashkovich2021,Wang2025}. While a wealth of spin-charge and spin-lattice coupling phenomena have been characterized in photo-excited antiferromagnets, these are often described within the framework of a three-temperature model that treats each subsystem as a distinct thermal reservoir \cite{Kirilyuk2010,Li2022Orthoferrites}. In contrast, nonthermal light-induced phase transitions that exhibit symmetry-breaking signatures from the direct, concerted interplay of charge, lattice, and spin remain rare and are of immense fundamental and technological interest.

Here, we employ time-resolved reflectivity and birefringence measurements to demonstrate an ultrafast laser-induced phase transition in $\alpha$-MnTe. Tracking the dynamics of the electronic, structural, and magnetic sectors reveals that all three degrees of freedom respond cooperatively cross the phase boundary at a single, well-defined fluence threshold.  The transition is initiated within the charge and spin subsystems -- manifesting as a rapid optical gap collapse and demagnetization -- before propagating to the lattice, which exhibits a clear symmetry-breaking signature in the form of an emergent side peak in the coherent phonon spectrum. Drawing on these observations, we provide mechanistic insights into the unique spin-charge and spin-lattice coupling pathways operative in this compound, and discuss their potential relevance to altermagnetism -- a recently identified magnetic class arising from an expanded spin-group classification framework for which $\alpha$-MnTe stands as a prototypical material \cite{ifmmodeSelseSfimejkal2022,Krempasky2024,Lee2024,GonzalezBetancourt2023,Yuan2020,Hu2025,Bossini2021,Plouff2025,Eskandariasl2025,Liu2023,Huang2024,Mazin2023,Zhu2023,Osumi2024,Kluczyk2024,Amin2024,Hubert2025,Liu2024,Hariki2024,Yang2026MnTeMOKE,Huang2026Photomodulated}.

\begin{figure}[t]
	\centering
	\includegraphics[width=\linewidth]{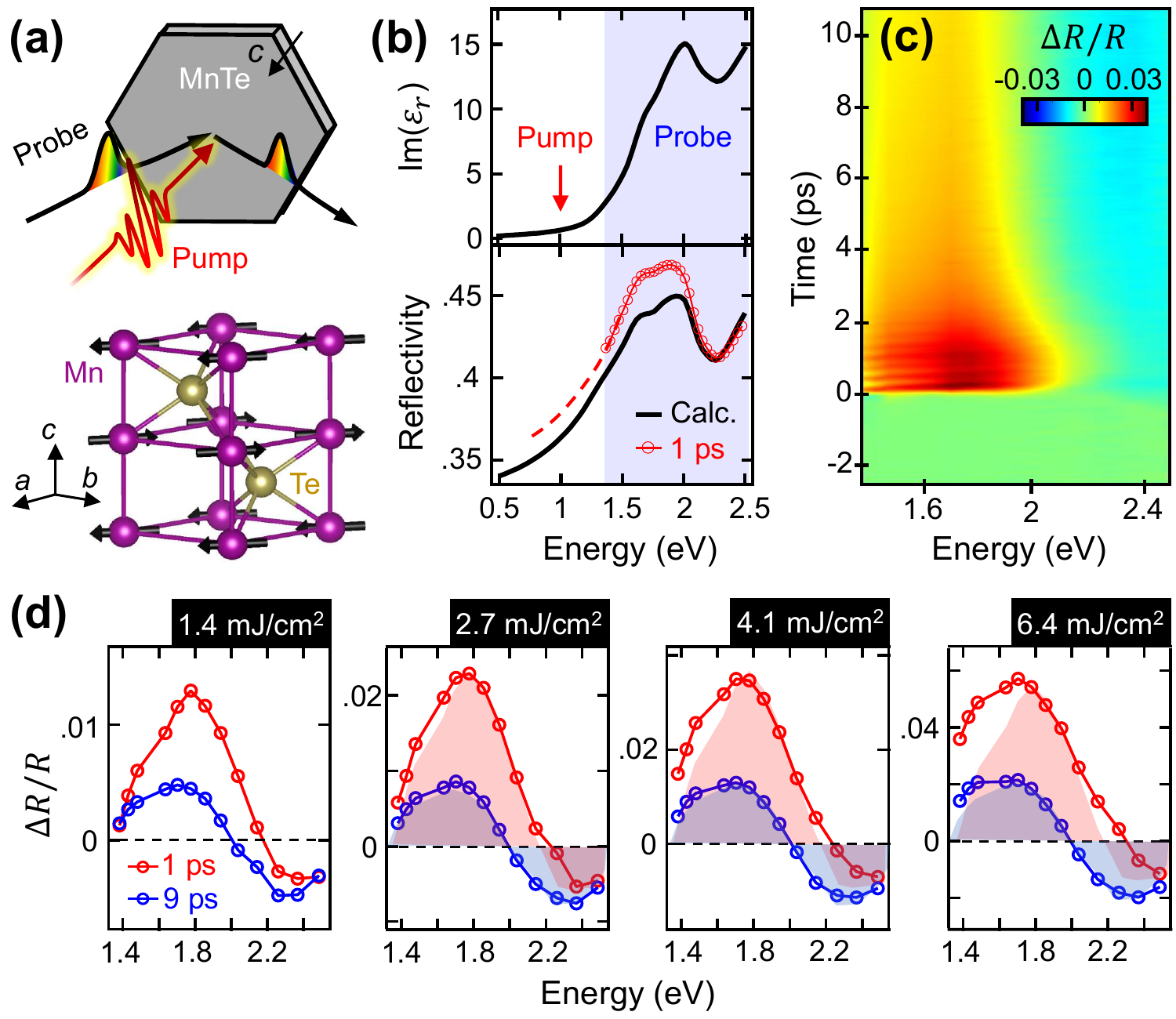}
	\caption{\small (a)~Pump-probe measurement geometry and crystal structure of MnTe. (b)~Calculated imaginary part of the relative permittivity and reflectivity spectrum, with nonequilibrium reflectivity at $t=1$~ps reconstructed from data (red line with markers) and its extrapolation (by red dashed line). (c)~$\Delta R/R$ mapped versus $t$ and $E$ at $F=4.1$~mJ/cm$^2$. (d)~$\Delta R/R$ spectra at $t=1$~ps and $t=9$~ps for various $F$, with data at $F=1.4$~mJ/cm$^2$ scaled and overlaid as shadings on higher-$F$ panels for comparison. All measurements here were performed at $T=173$~K.
	}
	\label{Fig1}
\end{figure}

$\alpha$-MnTe (MnTe) is an antiferromagnetic semiconductor with a hexagonal lattice composed of a face-sharing network of octahedra, where magnetic Mn atoms occupy the centers and Te atoms reside at the vertices. Upon magnetic ordering, the compound exhibits ferromagnetic spin alignment within the basal planes, which are coupled antiferromagnetically between adjacent layers [Fig.\,1(a)]. We first performed pump-probe reflectivity spectroscopy measurements to investigate the transient electronic dynamics [Fig.\,1(a)]. A freshly cleaved MnTe crystal ([0001] orientation) was excited by a near-infrared pump pulse (1~eV, 100~fs) and subsequently probed by a broadband white-light pulse (1.3--2.5~eV) at variable time delays. According to our calculated optical absorption spectrum, shown as the imaginary part of the relative permittivity $\varepsilon_r$ [Fig.\,\ref{Fig1}(b)], the 1 eV pump was resonant with the absorption gap edge primarily attributed to the charge-transfer excitations from Te $5p$ to Mn $3d$ orbitals \cite{SM}. The probe photon energy range was sensitive to reflectivity changes around and above the absorption gap.

Figure~1(c) presents the time-resolved differential reflectivity ($\Delta R/R$) spectra recorded at a temperature ($T$) of 173~K and a pump fluence ($F$) of 4.1~mJ/cm$^2$. At a time delay ($t$) of 1~ps following the pump excitation, the $\Delta R/R$ spectrum exhibits a pronounced positive peak at 1.8~eV, a negative dip at 2.4~eV, and a zero-crossing at 2.2~eV which shifts to lower energy with increasing $t$. Utilizing the calculated equilibrium reflectivity spectrum,  we reconstructed the reflectivity spectrum of the photo-excited state at this specific time snapshot [red curve in Fig.\,\ref{Fig1}(b)], which reveals a substantial increase in spectral weight at the 2 eV peak and in the low-energy region near the absorption edge, suggesting photo-induced gap shrinkage. Figure~1(d) displays the $\Delta R/R$ spectra at $t=1$~ps and $t=9$~ps, as well as their evolution with increasing $F$. While the overall trend is an increase in the differential signals with rising $F$, the low-energy region (1.4--1.6~eV) of the $t=1$~ps spectrum grows more rapidly, leading to noticeable spectral deformations at high $F$.

\begin{figure*}[thb]
	\centering
	\includegraphics[width=\linewidth]{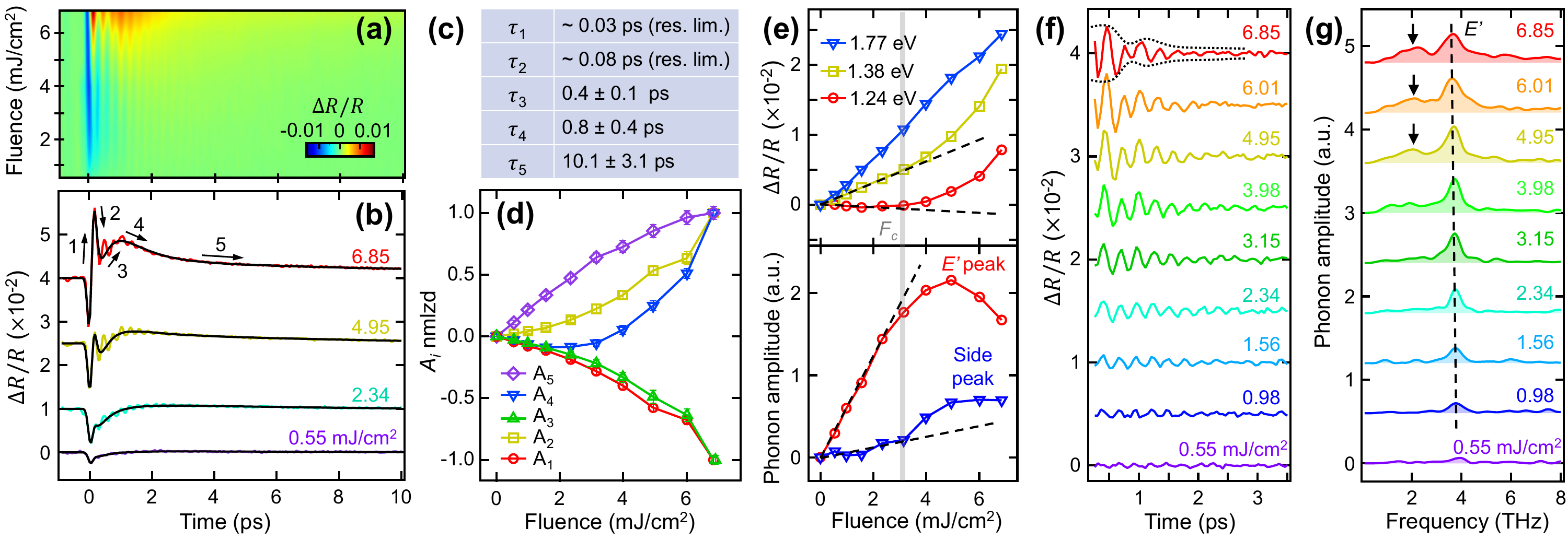}
	\caption{\small (a)~$\Delta R/R$ mapped versus $F$ and $t$, and (b)~time-domain cuts at selected $F$ (black lines: multi-exponential fits). (c)~Time constants $\tau_i$ and (d)~amplitudes $A_i$ extracted from fits to $\Delta R/R(t)$. (e)~$F$-dependence of $\Delta R/R$ at $t=1$~ps for three probe energies (upper panel), compared with the $F$-dependence of the $E'$ phonon amplitude (lower panel), revealing a common threshold fluence $F_c$. (f)~Background-subtracted $\Delta R/R$ transients showing coherent phonon oscillations and (g)~their Fourier transforms; curves are offset for clarity. All measurements here were performed at $T=153$~K and $E=1.24$~eV.
	}
	\label{Fig2}
\end{figure*}

To examine the possibility of a photo-induced electronic transition, we conducted detailed $F$-dependent measurements at a probe photon energy of $E=1.24$~eV while maintaining $T=153$~K. Figure~2(a) presents a colorplot of $\Delta R/R$ mapped as a function of $F$ and $t$, with time-domain cuts at selected $F$ shown in Fig.~2(b). Each $\Delta R/R$ trace generally features an electronic background superimposed with coherent-phonon-induced periodic modulations. Upon photoexcitation at time zero, $\Delta R/R$ drops negative, followed by complex dynamics that evolve through as many as five distinct steps at higher fluences (e.g., the $F=6.85$~mJ/cm$^2$ trace). We fitted the signal  using a convolution between a Gaussian instrumental resolution function and a five-component exponential decay $\sum_{i=1}^{5} A_i e^{-t/\tau_i}$, where  $A_i$ and $\tau_i$ are the amplitude and decay time of the $i$-th step, respectively. The extracted time constants $\tau_i$ are tabulated in Fig.~2(c): $\tau_1$ and $\tau_2$ approach the instrumental resolution (though $\tau_1<\tau_2$ remain distinguishable), $\tau_3$ and $\tau_4$ are fully resolved sub-picosecond dynamics, and $\tau_5$ is a slow step of 10.1~ps. Figure~2(d) plots the $A_i$ versus $F$, showing that all amplitudes except $A_5$ exhibit a superlinear response. In the raw data, this manifests as an accelerated reddening of the $\Delta R/R$ colormap at early time delays [Fig.~2(a)]. An $F$-dependent cut of the colormap at $t=1$~ps [red trace, upper panel of Fig.~2(e)] shows that $\Delta R/R$ varies linearly (with negative slope) up to $3.1$~mJ/cm$^2$ before a sharp nonlinear upturn above threshold. This behavior is also visible as a slope change at $E=1.38$~eV (brown) but absent at $E=1.77$~eV (blue). Together, these observations suggest that the photo-induced electronic transition drives an accelerated shrinkage of the band gap that majorly affects the low-energy region of the optical spectrum, manifesting primarily through steps 1--4 of the dynamics. We confirmed the nonthermal nature of this response by varying the pump photon energy and observed that $\Delta R/R$ changed sign, while the threshold behavior persisted \cite{SM}.

To gain insight into the lattice dynamics, we analyzed the coherent phonon oscillations superimposed on the electronic background. Figure~2(f) presents the background-subtracted $\Delta R/R$ transients at various $F$, with their fast Fourier transforms shown in Fig.~2(g). At the lowest $F$ of 0.55~mJ/cm$^2$, the spectrum exhibits a single peak at 3.75~THz -- a mode previously reported \cite{Zhang2020,Kluczyk2024,Gray2024,Wu2025}, but recently reassigned as the transverse-optical $E'$ phonon ($D_{3h}$ group) following the discovery of a native noncentrosymmetric structural distortion \cite{Wu2025}. 

%Notably, our analysis window is an order of magnitude shorter than that typical in picosecond ultrasonics \cite{Mattern2023}, where acoustic strain propagation---which does not affect optical phonons---is the primary focus.

Interestingly, the amplitude of the $E'$ peak does not exhibit a monotonic $F$ dependence. As shown by the red curve in the lower panel of Fig.\,2(e), the extracted peak amplitude first increases then decreases with $F$. Meanwhile, a new side peak emerges near 2~THz at large $F$ [Fig.\,2(g)]. This new peak, which has a linewidth (1~THz) smaller than its center frequency and thus qualifies as a distinct optical phonon mode, gives rise to a clear beating pattern in the time domain; see the $F=6.85$~mJ/cm$^2$ trace in Fig.~2(f). While strong optical excitation typically shifts or damps existing phonon modes \cite{Wall2012,Ning2023} -- often restoring higher structural symmetry -- emergence of an additional phonon mode at high fluence is a rare phenomenon  that points definitively to nonthermal laser-induced effects \cite{Li2019}. This behavior signals a photo‑driven lattice instability that lowers the crystal symmetry and lifts phonon degeneracy. Moreover, comparison across the upper and lower panels of Fig.\,2(e) reveals that the threshold fluence for the onset of phonon nonlinearity coincides with $F_c=3.1$~mJ/cm$^2$, matching that of the electronic transition.

Given that the electronic and structural degrees of freedom act in concert during the ultrafast transition, the key question concerns the response of the spin system. To investigate this, we employed time-resolved birefringence measurements to probe the dynamics of the antiferromagnetic magnetic order parameter (MOP). Figure\,3(a) illustrates the measurement principle. The formation of the MOP -- with $\pm \bm{M}$ representing the two opposite sublattices -- breaks the three-fold rotational symmetry along the $c$-axis that is otherwise preserved by the nonmagnetic crystal structure. This symmetry breaking induces in-plane optical anisotropy, yielding different refractive indices along ($n_1$) and perpendicular ($n_2$) to the N\'eel vector \cite{Bossini2021}, and thus birefringence $\delta n = n_1 - n_2$. When an $s$-polarized probe pulse is incident on the crystal, this birefringence generates a finite cross-polarized ($p$-polarized) component in the reflected light. We placed a polarizer (LP2) in the reflection path and measured the cross-polarized intensity, which scales as $I^{\perp}\propto\delta n^2/(n+1)^4$ \cite{SM}, where $n=(n_1+n_2)/2$. By capturing the pump-induced modulation $\Delta I^{\perp}=I^{\perp}_\text{pumped}-I^{\perp}_\text{unpumped}$ as in all time-resolved measurements, we obtain \cite{SM}
\begin{equation}
    \Delta I^{\perp}/I^{\perp}\propto 2\frac{\Delta(\delta n)}{\delta n} - 4\frac{\Delta n}{n+1},
\end{equation}
which establishes that our measured differential transient depends linearly on the pump-induced change in birefringence (first term) while also containing a photo-refractive contribution (second term).

\begin{figure*}[thb]
	\centering
	\includegraphics[width=\linewidth]{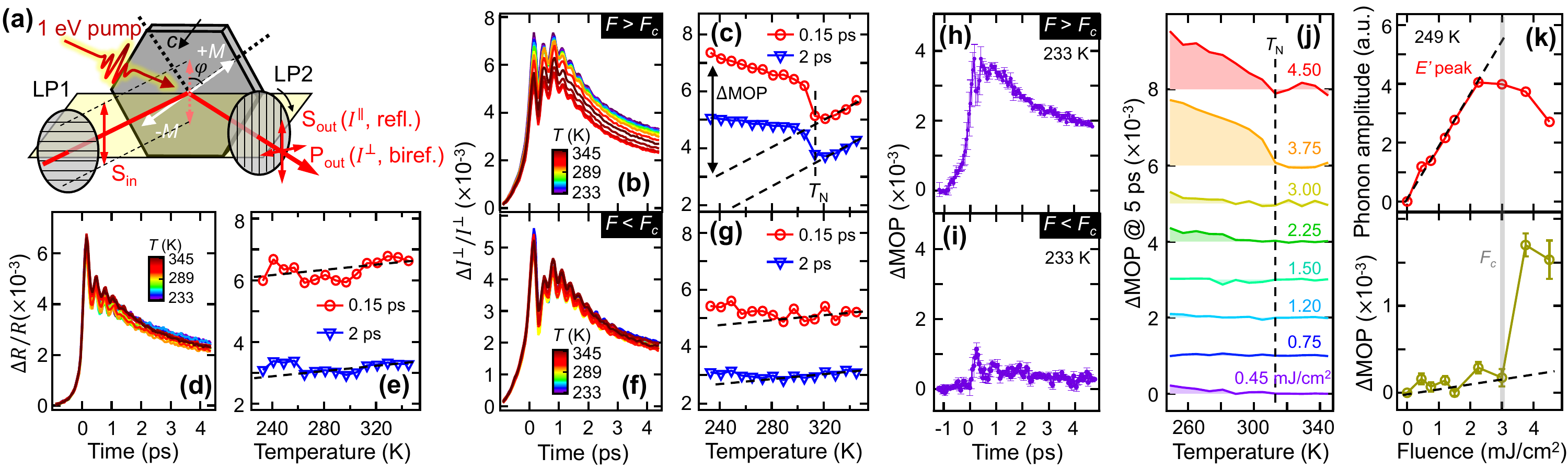}
	\caption{\small (a) Time-resolved birefringence measurement geometry. (b) $\Delta I^{\perp}/I^{\perp}$ transients at various $T$ ($F=3.75$~mJ/cm$^2>F_c$), and (c) their $T$-dependent cuts at two time delays; the black dashed lines mark the nonmagnetic backgrounds. (d,e) Same as (b,c) but for $\Delta R/R$ (co-polarized reflectivity). (f,g) Same as (b,c) but at $F=3$~mJ/cm$^2<F_c$. (h,i) $t$-dependent $\Delta$MOP at $T=233$~K, extracted from (b) and (f), respectively. (j) $T$-dependent $\Delta$MOP at various $F$ (background subtracted). (k) $F$-dependence of the $E'$ phonon amplitude compared with $\Delta$MOP at $t=5$~ps, revealing a common threshold fluence $F_c$ ($T=249$~K). The pre-time-zero slow rise in (b), (d), and (f) is due to a probe pulse artifact discussed in Ref.~\cite{SM}.%$E'$ and phonon sideband amplitudes are extracted from the 3.75-THz and 2-THz cuts in (e).
	}
	\label{Fig3}
\end{figure*}

Figure~3(b) displays the $t$-dependent $\Delta I^{\perp}/I^{\perp}$ at various $T$ ($F=3.75$~mJ/cm$^2$, $E=1.6$~eV); $T$-dependent cuts at $t=0.15$~ps and $t=2$~ps are shown in Fig.~3(c). Clear order-parameter-like onsets appear at the reported N\'eel temperature $T_\text{N}=310$~K. We believe this magnetic sensitivity originates from the first term in Eq.~(1), while the $T>T_\text{N}$ background (black dashed lines) arises from the nonmagnetic photo-refractive contribution. Supporting this assignment, our $\Delta R/R$ measurements (a direct probe of photo-refraction) under identical conditions -- obtained by rotating LP2 to capture co-polarized reflected light ($I^\parallel$) -- show no anomaly at $T_\text{N}$ [overlapping curves in Fig.~3(d) and no upturn in Fig.~3(e)]. In Fig.~3(c), we therefore attribute the difference between the $\Delta I^{\perp}/I^{\perp}$ curves and the nonmagnetic background to the pump-induced change in MOP, denoted $\Delta$MOP, which in this scenario corresponds to light-induced demagnetization of the antiferromagnet. We further measured $\Delta I^{\perp}/I^{\perp}$ under identical conditions but at a lower $F=3$~mJ/cm$^2$ [Fig.~3(f)]; the magnetic anomaly and $\Delta$MOP are greatly reduced [Fig.~3(g)]. The $t$-dependent $\Delta$MOP at 233~K, extracted from Figs.~3(b) and (f) via background subtraction, is plotted in Figs.~3(h) and (i), respectively. The former reveals an initial, resolution-limited demagnetization followed by slower evolution, while the latter shows little signal.

%This demagnetization is largely complete by $t=0.15$~ps, as shown by the extracted $t$-dependent $\Delta$MOP at $T=233$~K in the inset to Fig.~3(b), obtained by applying background removal at every time delay.

We further captured the full $F$ dependence of $\Delta I^{\perp}/I^{\perp}$. The resulting $\Delta$MOP traces at $t=5$~ps at various $F$ are summarized in Fig.~3(j). High-$F$ traces ($3.75$ and $4.5$~mJ/cm$^2$) exhibit large $\Delta$MOP, while lower-$F$ ones remain silent, suggesting a threshold fluence for photo-induced demagnetization reminiscent of the electronic and structural transitions. Remarkably, the threshold for the magnetic transition coincides with that for the structural one. This is confirmed by extracting the $E'$ phonon oscillations from the same dataset \cite{SM} and analyzing their amplitude's nonlinear $F$ dependence; Fig.~3(k) shows that the threshold $F_c$ for phonon nonlinearity coincides with the abrupt jump in $\Delta$MOP ($T=249$~K).

\begin{figure}[tbh]
	\centering
	\includegraphics[width=\linewidth]{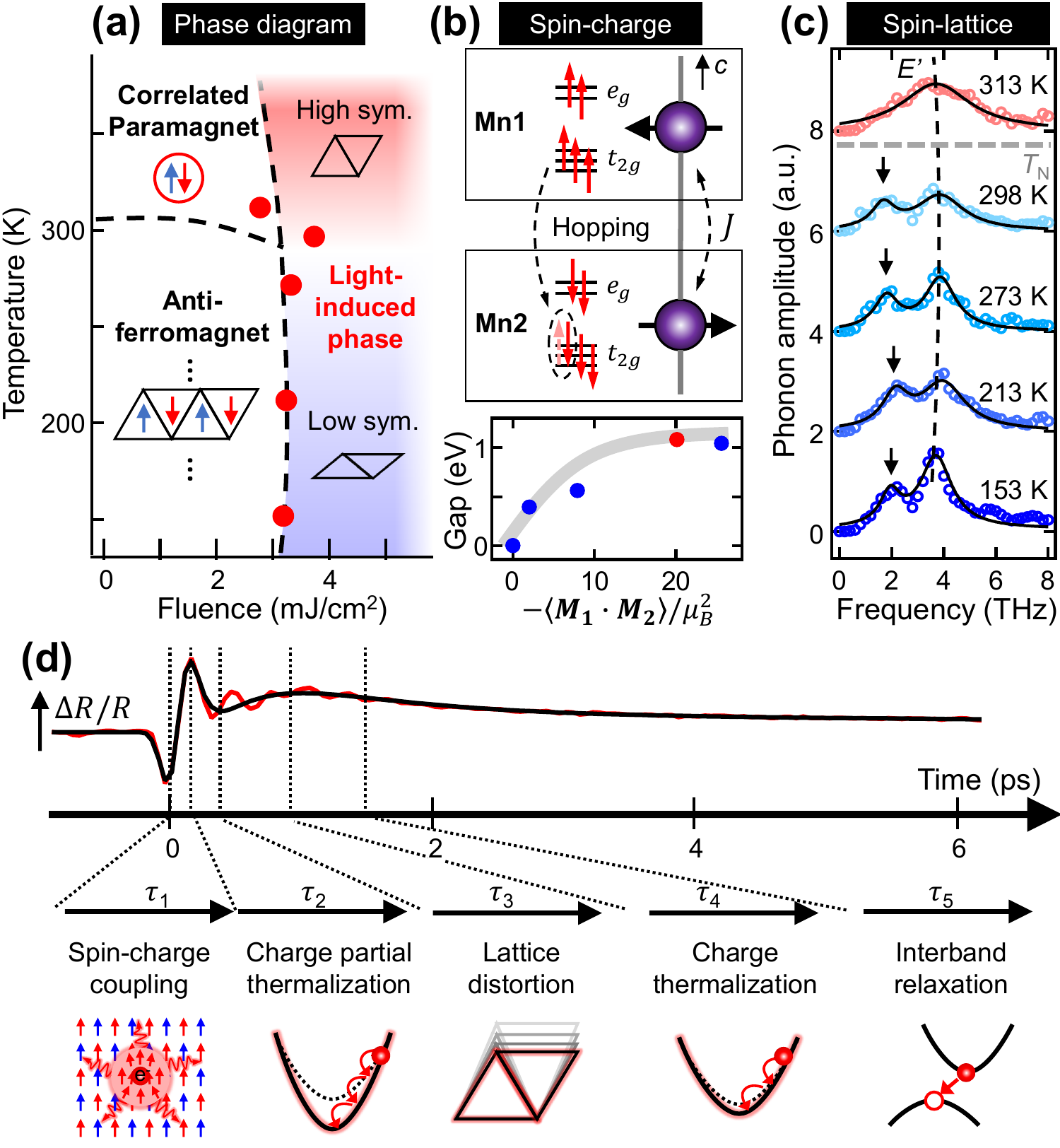}
	\caption{\small (a) $T$--$F$ phase diagram; red circles mark the experimentally determined phase boundary for the cooperative transition. (b) Upper panel: schematic of antiferromagnetic exchange between neighboring Mn atoms along the $c$-axis. Lower panel: calculated optical gap size versus magnetic correlation strength; the red circle indicates the equilibrium condition. (c) High-fluence ($F>F_c$) coherent phonon spectra at selected temperatures, fitted with double-Lorentzian (below $T_\text{N}$) or single-Lorentzian (above $T_\text{N}$) profiles (black lines). The $T=153$~K spectrum is the same as the $F=6.85$~mJ/cm$^2$ curve in Fig.~2(g). (d) Schematic of the proposed multistage temporal evolution following photoexcitation at $t=0$. 
	}
	\label{Fig4}
\end{figure}

Through an extensive exploration of the parameter space defined by $T$ and $F$, we found that magnetic transition always occurs in synchrony with structural and electronic transition. Accordingly, a single phase boundary can be established for the cooperative photo-induced electronic, structural, and magnetic transition. The experimentally determined phase boundary is marked by red circles in the $T$--$F$ phase diagram in Fig.\,4(a).  This boundary extends even above $T_\text{N}$, where clear signatures of photo-induced electronic and phonon nonlinearity persist despite the absence of magnetic order \cite{SM}.

Below, we discuss the key roles of spin-charge and spin-lattice coupling in the cooperative transition. First, the observed optical-gap shrinkage can be understood as a back-action of photo-induced demagnetization on the electronic sector. Stabilized by half-filled $d$ orbitals of Mn$^{2+}$, MnTe exhibits strong nearest-neighbor antiferromagnetic exchange along the $c$-axis [Fig.~4(b)], sustaining short-range spin correlations well above $T_{\text{N}}$ \cite{Baral2022}. When a pump pulse generates interband photo-charges, these correlations are nonthermally quenched via changes in orbital occupation -- a process we modeled using a two-site approach \cite{SM}. The melting of short-range correlations (both below and above $T_\text{N}$) then back-acts on the charge sector, driving accelerated gap shrinkage. Our \textit{ab initio} calculations confirm that the charge gap decreases as magnetic correlation weakens [lower panel of Fig.~4(b)] \cite{SM,Bossini2020}. This picture naturally explains why the phase boundary extends above $T_\text{N}$ and why low-$F$ pumping slightly below $T_\text{N}$ fails to substantially demagnetize [Fig.~3(j)]: in both cases, the magnetic correlations remain robust against weak perturbation.

Second, the structural dynamics further underscore the role of spin-lattice coupling. As shown in Fig.~4(c), the new phonon peak emerging for $F>F_c$ -- a signature of lattice symmetry breaking -- appears only for $T<T_\text{N}$, indicating that photo-induced symmetry lowering occurs exclusively when the initial phase is long-range ordered, an effect unseen in classical magnets \cite{KorffSchmising2008,Reid2018,Pudell2019}. To interpret this, we invoke a possible mechanism rooted in altermagnetic piezomagnetism \cite{Aoyama2024,McClarty2024}. For MnTe, a $g$-wave altermagnet, theory predicts that ultrafast photo-doping induces uncompensated demagnetization between sublattices \cite{Zhou2026CrSb}: the staggered moment diminishes unevenly, producing transient net magnetization $\bm{\mathcal{M}}$ within tens of fs after photo-excitation. This enables symmetry-breaking stress $\sigma_{ij} = \Lambda_{ijk} \mathcal{M}_k$, where $\Lambda_{ijk}$ is a rank-three axial $c$-type tensor that vanishes in conventional antiferromagnets but is finite in altermagnets due to broken time-reversal and parity-time symmetries. The resulting stress can be inherently short-lived given the rapid demagnetization observed, but a sufficiently strong impulsive kick could impart enough momentum to drive substantial lattice distortion at later time delays.

Collectively, our results reveal the following temporal evolution following initial photoexcitation at $t=0$ [Fig.~4(d)]. (i) Within $\tau_1$: $\Delta R/R$ near the gap exhibits a strong positive upturn, signaling rapid gap shrinkage driven by ultrafast demagnetization. Concurrently, symmetry-breaking lattice stress begins to build up. (ii) Within $\tau_2$: $\Delta R/R$ partially relaxes toward equilibrium, indicating partial charge thermalization, presumably dominated by charge-charge scattering in the demagnetized high-symmetry lattice. (iii) Within $\tau_3$: Lattice strain accumulates in response to the transient stress from stage (i). The symmetry distortion fully develops by $\sim$0.75~ps, consistent with typical phonon timescales, producing the beating pattern in the coherent phonon oscillations [Fig.~2(f)], and drives a second upturn in $\Delta R/R$. (iv) Within $\tau_4$: Charge thermalization proceeds within the distorted lattice potential. (v) Within $\tau_5$: Interband charge recombination occurs. This is the only step lacking nonlinear threshold behavior and is thus prevalent even at low fluences [Fig.~2(b)].

Our demonstration opens an avenue for dynamical spin manipulation by harnessing coupled fundamental degrees of freedom -- via photo-magnetic \cite{Kalashnikova2015}, phono-magnetic \cite{Juraschek2020}, or coherent Floquet switching \cite{Chaudhary2019}. The altermagnetic piezomagnetic mechanism proposed here warrants further verification using time-resolved x-ray/electron diffraction and inelastic x-ray scattering. More broadly, the coupling pathways we observe find their conceptual counterparts in altermagnetic phenomena such as spin-selective optical excitations and hybrid magnon-phonon modes \cite{Vila2025,Steward2023,Haag2026BulkGwave}. These effects are rooted in symmetry principles that couple spin, charge, and lattice -- principles that trace back to key concepts proposed early on for the entire altermagnet class, including nonrelativistic spin splitting of bands and spin-group operations that entangle spin with sublattice. Looking ahead to the growing interest in optical probing and photo-manipulation of altermagnets, we envision that these pathways can be extended to the wider class of material candidates, encompassing both $d$-wave (e.g., MnF$_2$) and $g$-wave (e.g., CrSb) systems, paving the way for altermagnet-based opto-spintronic applications.

\par\smallskip\noindent\textit{Acknowledgments}---X.L. and C.L. acknowledge support from the SUSTech-NUS Joint Research Program (A-8002266-00-00). X.L. acknowledges support from the Singapore National Research Foundation under award number NRF-NRFF16-2024-0008 and the Ministry of Education, Singapore, Academic Research Fund Tier 2 under grant No. MOE-T2EP50124-0005. Work at SUSTech was supported by the National Key R\&D Program of China (No. 2022YFA1403700) and the National Natural Science Foundation of China (NSFC) (No. 12534003).

\par\smallskip\noindent\textit{Data availability}---The data supporting the findings of this study are available in Zenodo \cite{MnTeData}.

\end{document}